\documentclass[12pt]{article}
\usepackage{epsfig}

\title{Light Quark Mass Difference and Isospin
        Breaking In Electromagnetic Pion Production\footnote{to be 
published in Physics Letters B}}
\author{A.M.~Bernstein\\
        Physics Department and Laboratory for Nuclear Science\\
        M.I.T., Cambridge Mass., USA}

\begin{document}

\maketitle

\begin{abstract}
It is demonstrated that there is a dynamic isospin breaking effect in
the near threshold $\gamma^{*} N\rightarrow \pi N$ reaction due to 
the
mass difference of the up and down quarks, which also causes 
isospin
breaking in the $\pi N$ system. The photopion reaction is affected
through final state $\pi N$ interactions (formally implemented by
unitarity and time reversal invariance). It is also demonstrated that
the near threshold $\gamma \vec{N} \rightarrow \pi N$ reaction is a
practical reaction to measure isospin breaking in the $\pi N$ system,
which was first predicted by Weinberg about 20 years ago but has 
never
been experimentally tested.
\end{abstract}

Since the discovery of a large mass difference between the up and 
down
quarks ($m_{u}\simeq 5 MeV, m_{d} \simeq 9 MeV$) there has been
considerable interest in the possibility of observing dynamical
isospin breaking in the pion-nucleon system
\cite{W1,GL1,L,vanKolk,Meissner}. The theoretical consensus is that
the magnitude of isospin breaking is not $(m_{d}-
m_{u})/(m_{d}+m_{u})
\simeq 27 \% $, but instead is $\simeq
(m_{d}-m_{u})/\Lambda_{qcd}\simeq 2 \%$ \cite{L}, where 
$\Lambda_{qcd}
\simeq$ 0.2 GeV. In this paper it will be shown for the first time
that the $\gamma p\rightarrow \pi^{0}p$ reaction is an excellent
candidate to measure dynamic isospin violations due to the up, down
quark mass difference.

Weinberg first showed that there is an isospin violating effect in the
s wave $\pi N$ scattering length $a(\pi N)$ due to the up, down 
quark
mass difference \cite{W1}. This predicted effect, which occurs for
$\pi^{0}N$ scattering or charge exchange but not for $\pi^{\pm}N$
scattering, has more recently been calculated by Chiral Perturbation
Theory (ChPT) \cite{vanKolk, Meissner}. The predicted magnitude of
this effect is the same (to within a factor of $\sqrt{2}) $ in
$\pi^{0} N$ scattering and charge exchange reactions. However, since
the magnitude of $a(\pi^{0} N)$ is small, the relative magnitude of
the isospin violating term is $\simeq$ 30 \%
\cite{W1,vanKolk}.  By contrast, for charge exchange where
the isospin conserving amplitude is larger, the relative isospin
violation is estimated to be $\simeq$ 2 to 3 \% \cite{W1,vanKolk}.

In this paper the (dynamic) isospin breaking effect of the up and 
down
quark mass difference is shown to be present in the near threshold
$\gamma$ N $\rightarrow \pi$N reaction. This is in addition to the
well known (static) isospin breaking effect due to the threshold
difference between the $\gamma p \rightarrow \pi^{0}p$ and 
$\gamma p
\rightarrow \pi^{+}n$ channels. Of this $\simeq$ 2/3 is due to the
Coulomb effect and $\simeq$ 1/3 is caused by the up, down quark 
mass difference. The separation of the two threshold
energies leads to the prediction of a unitary cusp in the $\gamma p
\rightarrow \pi^{0}p$ reaction near the $\pi^{+}n$ threshold due to
the two step $\gamma p \rightarrow \pi^{+}n\rightarrow\pi^{0}p$
reaction \cite{BKM,cusp,AB,Berg}. The magnitude of this unitary cusp
is due to the large ratio of the electric dipole multipole amplitudes,
$R = E_{0+}(\gamma p\rightarrow \pi^{+}n)/ E_{0+}(\gamma 
p\rightarrow
\pi^{0}p)\simeq-20 $ \cite{BKM}. The first derivations of the unitary
cusp \cite{cusp} used a K matrix approach to calculate the effect of
charge exchange in the final state. Due to their single scattering
approximation, these calculations violate unitarity as will be
discussed below. The chiral perturbation theory (ChPT) calculations
are basically isospin conserving but the biggest isospin
non-conserving effect due to the pion mass difference (which is 
almost
purely electromagnetic in origin) is inserted by hand
\cite{BKM}. These calculations also violate unitarity due to their
truncation at the one loop (single rescattering) level. There has been
recent progress in performing ChPT calculations in $\pi N$ scattering
which take the isospin violations due to the Coulomb interaction into
account \cite{Meissner}.

The approximations made in the K matrix \cite{cusp} and published 
ChPT
papers \cite{BKM} can be overcome by using a 3 channel S matrix
approach in which unitarity and time reversal invariance are 
satisfied
(a preliminary version of this work has been presented previously
\cite{AB}). This has the advantage that both static and dynamic
isospin breaking, and final state multiple scattering (to all orders),
are taken into account. The S matrix for the 3 open channels 
$(\gamma
p, \pi^{0}p, \pi^{+}n)$ can be written as :

\begin{equation}
\left(
\begin{array}{lcr}
S_{\gamma}& i M'_{0}& iM'_{c}\\ iM'_{0}&S_{0}&S_{0c}\\
iM'_{c}&S_{c0}&S_{c}
\end{array}
\right)\label{eq1}
\end{equation}
where $S_{\gamma}=e^{2i\delta_{\gamma}}, S_{0}=\cos\phi
e^{2i\delta_{0}},S_{c}=\cos\phi e^{2i\delta_{c}}$ represent elastic
$\gamma p$, $\pi^{0}p$, $\pi^{+}n$ scattering with phase shifts
$\delta_{\gamma},\delta_{0}$, and $\delta_{c}$ respectively,
$S_{0c}=S_{c0}=i\sin\phi e^{i(\delta_{0}+\delta_{c})}$ represents the
$\pi^{+}n\leftrightarrow\pi^{0}p$ charge exchange amplitude, $\phi$
is a real number, and $\cos\phi$ represents the inelasticity due to
charge exchange. For convenience, the off diagonal matrix elements
for the photo-production of the $\pi^{0}p$ and $\pi^{+}n$ channels
are written as $i M'_{0}$ and $i M'_{c}$, where $M'_{0}$ and $
M'_{c}$ are proportional to the multipole amplitudes.  For the
important case of production of s wave pions these are the $E_{0+}
(L_{0+})$ transverse (longitudinal) multipoles.  Although not
explicitly written here, the S matrix elements are for a fixed value
of W, the total CM energy, and the quantum numbers l and j, the $\pi
N$ orbital and total angular momenta. The notation of using 0 (c) for
the neutral (charged) channel is conveniently generalized to neutron
targets where the three channels are $\gamma n, \pi^{0}n$, and
$\pi^{-}p$.

Time reversal invariance requires that the S matrix be symmetric
($S_{ij} = S_{ji}$) and unitarity requires that $S^{+}S = S S^{+} =
1$. The form of the 2x2 $\pi N$ portion of the S matrix has been
chosen to be separately unitary and time reversal
invariant. Eq.~\ref{eq1} is symmetric by construction. Applying the
unitary constraint and assuming the weakness of the electromagnetic
interaction by dropping terms of order e$^2$, one obtains \cite{Low}:

\begin{equation}
\begin{array}{rl}
 M'_{0} & = e^{i(\delta_{\gamma}+\delta_{0})}  [A'_{0} \cos
\frac{\phi}{2} + i A'_{c} \sin \frac{\phi}{2} ] \\
  M'_{c}   & =  e^{i(\delta_{\gamma}+\delta_{c})} [A'_{c} \cos
\frac{\phi}{2}
+ i A'_{0} \sin \frac{\phi}{2} ]
\end{array}\label{eq2}
\end{equation}
where $A'_{0}$ and $A'_{c}$ are smoothly varying, real functions, of
the CM energy, and can be identified as proportional to the multipole
matrix elements $M_{0} $(for $\gamma p \rightarrow \pi^{0}p$) and
$M_{c}$ (for $\gamma p \rightarrow \pi^{+}n$) in the absence of the
final state $\pi N$ interaction. These equations show the connection
between electromagnetic pion production and $\pi N$ interactions. 
Note
that the phase shifts $\delta_{\gamma},\delta_{0}$, and $\delta_{c}$
appear as $e^{i \delta}$ in photopion reactions compared to $e^{2 i 
\delta}$ in elastic scattering because
the $\pi N (\gamma N)$ interactions take place in the final (initial)
state only. Eqs.~\ref{eq1} and \ref{eq2} are valid for both photo- and
electro-production, i.e. for both real and virtual photons.  In
general $M'_{0}, M'_{c}, A'_{0}$, and $A'_{c}$ are functions of
$q^{2}$ (the invariant four momentum transfer) as well as W, l, and
j. To order $e^{2}$ the $\pi N$ sector parameters
$\delta_{0},\delta_{c}$, and $\phi$ are functions of W, l, and j only.

In this treatment terms of order $e^{2}$ have been neglected with 
the
exception of $\delta_{\gamma}$ which is small compared to both
$\delta_{0}$ and $\delta_{c}$ (except at the $\pi^{0}$ threshold), and
the mass difference between the charged and neutral pions which is 
put
in by hand. This is the largest $O(e^{2})$ effect found in $\pi N$
scattering \cite{Meissner} and in photoproduction \cite{BKM}. The 
effect
of the $O(e^{2})$ terms have been worked out, although not 
presented
here. In general they produce a small coupling between the
electromagnetic and $\pi N$ sectors, so that e.g. the 2x2 $\pi N$
sector of the S matrix is no longer separately unitary. When the
experimental situation reaches a sufficient level of accuracy these
corrections should then be implemented.

Eq.~\ref{eq2} is a generalization of the final state interaction
theorem of Fermi and Watson \cite{FW} who first pointed out the
connection between photo-pion production and $\pi N$ scattering. In
their derivation, as in this one, time reversal and unitarity (to
order $e^{2}$) were assumed. However they made the additional
assumption that if one could neglect $e^{2}$ (and higher order 
terms)
isospin would be conserved (this was before the time of quarks). This
reduces the dimensionality of the S matrix to 2x2 which gives the
simple solution $S(\gamma p \rightarrow \pi N:I) = \pm| S |
e^{i\delta(\pi N:I)}$ where I=1/2,3/2 is the isospin of the $\pi N$
system. If isospin were conserved, so that the two thresholds are
degenerate, Eq.~\ref{eq2} reduces to the Fermi-Watson theorem. For 
the
near threshold region in which the CM pion momentum q 
$\rightarrow 0$,
the s wave phase shifts are $\delta_{0} = [2 a_{3} + a_{1}] q/3
=a_{\pi^{0}p} q, \delta_{c} = [a_{3}+2a_{1}]q/3= a_{\pi^{+}n} q$, and
$\phi/2 = [a_{1}-a_{3}] q/3 = a_{cex}q$ where $a_{i}$ is the s wave
$\pi N$ scattering lengths in the isospin 2I = 1, 3, or designated
charge states.  However isospin conservation is badly violated in the
threshold region due to the threshold energy difference between the
$\pi^{+} n$ and $\pi^{0} p$ channels and the additional dynamic
isospin violating effect due to the up and down quark mass 
difference.
This generalization of the Fermi-Watson theorem removes the
approximation of isospin conservation. Note the interesting feature
that below the $\pi^{+}n$ threshold there are only two open channels
and Eq.~\ref{eq2} reduces to the form of the Fermi-Watson theorem,
i.e. $M'_{0}$ = real number $\cdot\,e^{i \delta_{0}}$ or, equivalently,
that $tan(\delta_{0}) = \hbox{Im} E_{0+}/\hbox{Re} E_{0+}$.

It is of interest to show the connection between the S matrix
formulation which takes the final state scattering into account to all
orders and the original K matrix derivations \cite{cusp} which only
took a single final state scattering into account. This single
scattering limit can be taken by expanding $e^{i \delta_{0}}\simeq 1 
+
i \delta_{0}$, and neglecting second order terms like $\delta_{0}
\delta_{c}$, thus obtaining $M_{0} \simeq A_{0} + i A_{0} \delta_{0}
+i A_{c} a_{cex} q_{c}$. By examing the region below the $\pi^{+}n$
threshold where $q_{c}$ is continued as $i | q_{c} |$, it is observed
that $tan(\delta_{0}) \simeq \delta_{0} \neq \hbox{Im}
E_{0+}/\hbox{Re} E_{0+}$ This means that the single scattering
approximation does not obey unitarity, although in this particular
case the numerical effects are not large. The ChPT calculations to one
loop have a similar unitarity problem as was discussed by the ChPT
authors \cite{BKM}.

The connection between $M'_{0(c)}$ to the multipole amplitudes
$M_{0(c)}$ is $M'_{0(c)} = 2 \sqrt{ q_{\gamma} q_{0(c)}} M_{0(c)}$.
Note that the multipoles (M) have the dimensions of length, while the
S matrix elements (M') are dimensionless. The s wave 
photoproduction
cross section is $\sigma_{0(c)}(\theta) = | M'_{0(c)}|^{2}/4
q_{\gamma}^{2} = (q_{0(c)}/q_{\gamma}) | M_{0(c)}
|^{2}$. Similarly the s wave $\pi^{+}n \rightarrow \pi^{0}p$ charge
exchange reaction cross section is $\sigma_{cex}(\theta) =
\sin^{2}\phi/4q_{c}^2 = (q_{0}/q_{c}) | f_{cex} |^{2}$ from
which one obtains $\sin\phi = 2 \sqrt{q_{0} q_{c}} | f_{cex} |$
\cite{notation}. Defining $A'_{0(c)} = 2 \sqrt{q_{\gamma}
q_{0(c)}}A_{0(c)}$ one can then write Eq.~\ref{eq2} in terms of
multipole amplitudes as:

\begin{equation}
\begin{array}{rl}
 M_{0} & = e^{i(\delta_{\gamma}+\delta_{0})}  [A_{0} \cos
\frac{\phi}{2} + i A_{c} q_{c} | f_{cex} |/cos \frac{\phi}{2} ]
\\
  M_{c}   & =  e^{i(\delta_{\gamma}+\delta_{c})} [A_{c} \cos
\frac{\phi}{2}
+ i A_{0} q_{0} | f_{cex} |/\cos \frac{\phi}{2} ]
\end{array}\label{eq3}
\end{equation}

In the threshold region Eqs.~\ref{eq2} and \ref{eq3} lead to a unitary
cusp. In this case, $\phi \rightarrow a_{cex} q_{c},
\cos(\frac{\phi}{2}) \rightarrow 1, | f_{cex}| \rightarrow a_{cex}$,
and $A_{c} \simeq E_{0+}(\gamma p \rightarrow \pi^{+}n)$ for the
transverse multipole, or $L_{0+}(\gamma p\rightarrow \pi^{+}n))$ for
the longitudinal multipole. Below the $\pi^{+}n$ threshold $q_{c}$ is
continued as $i | q_{c}|$. Thus the contribution of the two step
$\gamma p \rightarrow \pi^{+}n \rightarrow \pi^{0}p$ reaction is in
$\hbox{Re} M_{0} (\hbox{Im} E_{0+})$ below (above) the $\pi^{+}n$
threshold. The magnitude of the unitary cusp, $\beta$, is defined as
the coefficient of $q_{c}$:

\begin{equation}
\beta= A_{c} \left| f_{cex}(\pi^{+}n\rightarrow\pi^{0} p)\right|
/cos(\phi/2) \simeq E_{0+}(\gamma p \rightarrow\pi^{+}n)
a_{cex}(\pi^{+}n\rightarrow\pi^{0} p)\label{eq4}
\end{equation}
The value of $\beta$ can be calculated from the experimental value 
of
$a_{cex}(\pi^{-}p \rightarrow \pi^{0}n)$ \cite{PSI}.  Assuming isospin
is conserved $a(\pi^{+}n \leftrightarrow \pi^{0}p) = -a(\pi^{-}p
\leftrightarrow \pi^{0}n)$ \cite{sign}. For $E_{0+}(\gamma p
\rightarrow\pi^{+}n)$ the ChPT prediction of $28.2\pm 0.6$
\cite{units,BKM} is used. This is in agreement with the preliminary
value of $27.6\pm 0.3$ from a recent experiment
\cite{Korkmaz,units}. Combining the two experimental (theoretical)
numbers gives $\beta= 3.59\pm 0.17$ $(\beta= 3.51\pm 0.22)$
\cite{units}.

Although it has not been previously noticed, the same physical
mechanism that causes the isospin breaking in $\pi N$ scattering and
charge exchange also enters into $\beta$. This follows from unitarity
and time reversal (Eqs.~\ref{eq2} and \ref{eq3}), combined with the
prediction of isospin breaking for the s wave scattering lengths for
charge exchange reactions $\delta a_{cex}$, that there is an isospin
violating contribution to $\beta$,

\begin{equation}
\begin{array}{rl}
\delta\beta= E_{0+}(\gamma p \rightarrow\pi^{+}n) \delta
a_{cex}(\pi^{+}n\rightarrow\pi^{0} p) \\ \delta\beta/\beta = \delta
a_{cex}(\pi^{+}n\rightarrow\pi^{0} p)/
a_{cex}(\pi^{+}n\rightarrow\pi^{0} p)
\end{array}\label{eq5}
\end{equation}
In the second line it has been assumed that there is no quark mass
effect on $E_{0+}(\gamma p \rightarrow \pi^{+}n)$. It is estimated
that $\delta a_{cex}/a_{cex} \simeq$ 2 to
3\% \cite{W1,vanKolk}. A better theoretical calculation of this effect
should be performed.

\begin{figure}
\begin{center}
\epsfig{figure=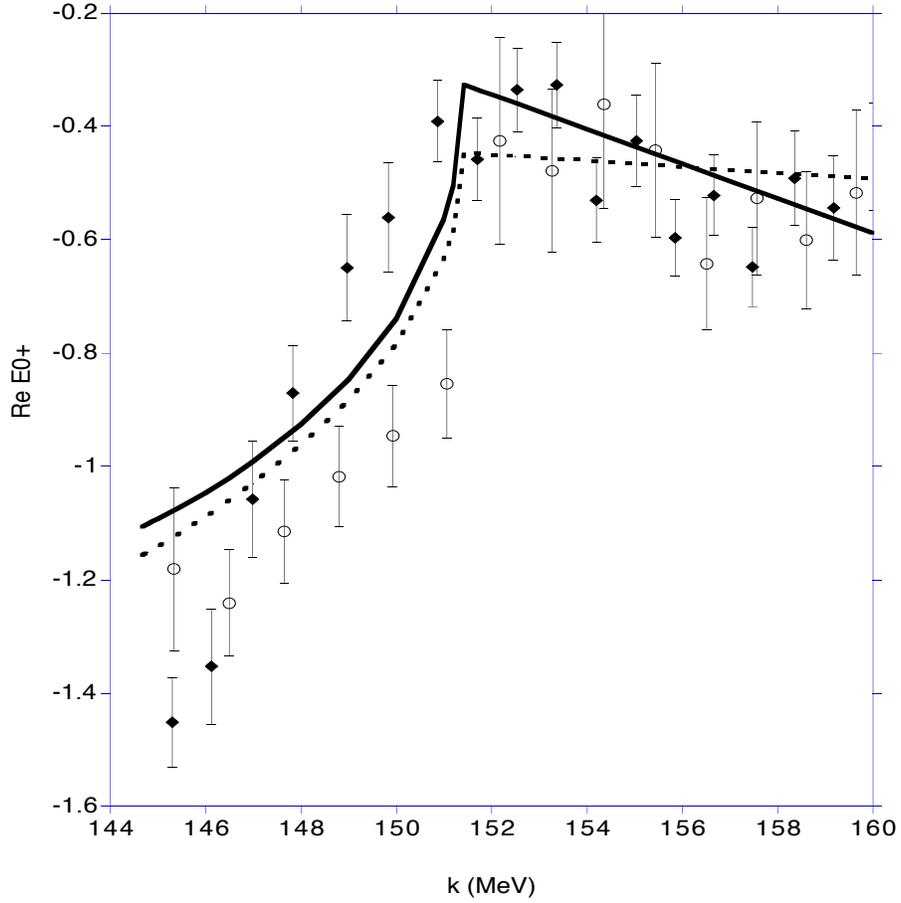,width=12cm,height=12cm}
\end{center}
\caption{$\hbox{Re} E_{0+}$ (in units of $10^{-3}/m_{\pi}$) for the
$\gamma p \rightarrow \pi^{0}p$ reaction versus photon energy k.  
The
dashed dot curve is the ChPT fit \cite{BKM} and the solid curve is the
unitary fit (Eq.~\ref{eq1}) \cite{Mainz}. The solid diamonds (circles)
are the Mainz (Saskatoon) points \cite{Mainz,Sask}.  The errors are
statistical only.}\label{fig1}
\end{figure}

\begin{figure}
\begin{center}
\epsfig{figure=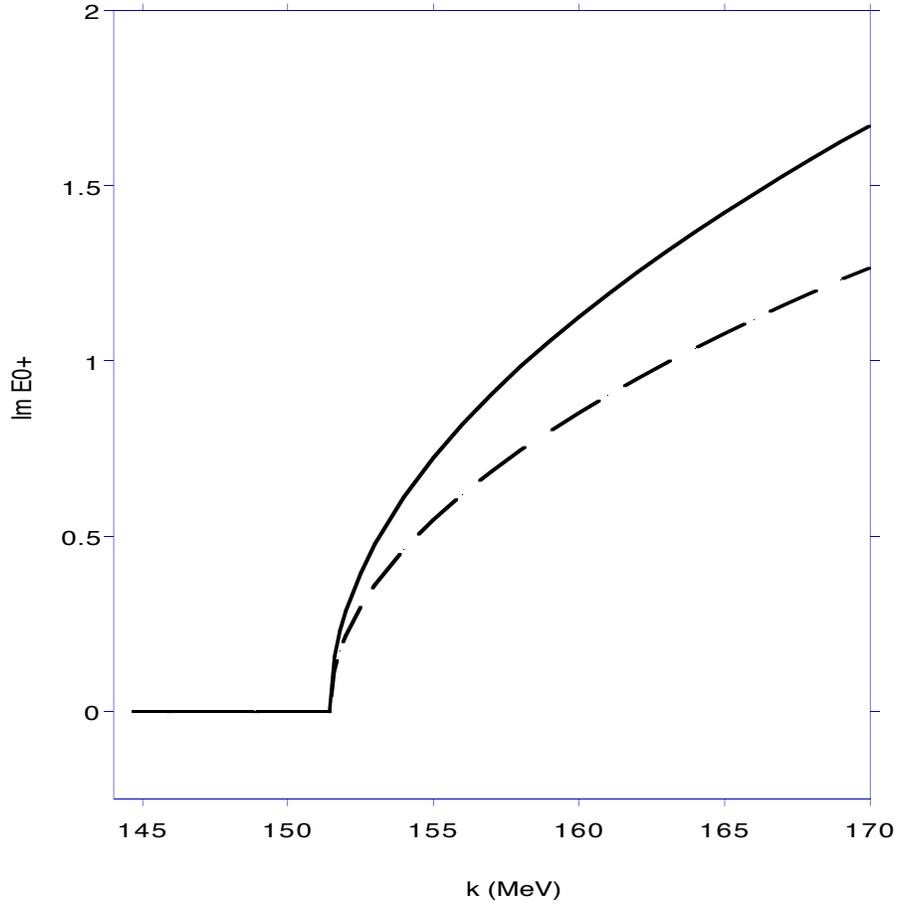,width=12cm,height=12cm}
\end{center}
\caption{$\hbox{Im} E_{0+}$ (in units
of $10^{-3}/m_{\pi}$) for the
$\gamma p \rightarrow \pi^{0}p$ reaction versus photon energy k.
The dashed dot curve is the ChPT calculation \cite{BKM} and the solid
curve is the unitary calculation (Eq.~\ref{eq1}).}\label{fig2}
\end{figure}

The electric dipole amplitude $E_{0+}(\gamma p \rightarrow 
\pi^{0}p)$
with its unitary cusp is presented in Figs.~\ref{fig1} and
\ref{fig2}. The recent Mainz/TAPS \cite{Mainz} and Saskatoon
\cite{Sask} results for $\hbox{Re} E_{0+}$ are presented in
Fig.~\ref{fig1}, where only the statistical errors are shown. The
small deviations between the data sets suggest the magnitude of the
systematic errors. Considering this, the agreement between the
calculated curves and the data is satisfactory. The curves are the
ChPT calculation with three empirical low energy parameters used in
fitting the data \cite{BKM}, and a unitary fit to the data
\cite{Mainz} using Eq.~\ref{eq3} which has $\beta=3.67$ \cite{units}
and a linear function of photon energy for $A_{0}$ with two 
parameters
which were fit to the data. For this case $R = E_{0+}(\gamma p
\rightarrow\pi^{+}n)/ E_{0+}(\gamma p \rightarrow \pi^{0}p)\simeq 
- 20
$ \cite{BKM} and the effect of the two step charge exchange reaction
is dramatic in the real (imaginary) part below(above) the $\pi^{+}n$
threshold.  Approximately 1 MeV above the $\pi^{+}n$ threshold the
$|\hbox{Im} E_{0+}| = |\hbox{Re} E_{0+}|$. The negative sign
in $R$ makes $\hbox{Im} E_{0+}$ have the opposite sign from 
$\hbox{Re}
E_{0+}$.

It is interesting to note that both the power counting rules of ChPT 
and the constraints of unitarity  lead to pion rescattering in the
final state as a critical dynamical ingredient. Although the unitary
and ChPT curves both agree with the data for $\hbox{Re} E_{0+}$ 
there
is an important difference between them. The value of $\beta$ = 
2.78
\cite{BKM, units} calculated for the $\gamma p \rightarrow \pi^{0}p$
reaction is smaller then the one calculated using the separately
predicted values of $E_{0+}(\gamma p \rightarrow\pi^{+}n)$ and $
a_{cex}(\pi^{+}n\rightarrow\pi^{0}p)$ of $3.51\pm 0.22$. This
difference can be clearly seen in Fig.~\ref{fig2}. The reason for this
discrepancy, which was discussed by the ChPT authors \cite{BKM}, is
due to the fact that the ChPT calculation is carried out to one loop
which is not sufficient for $\hbox{Im} E_{0+}$. This is a general
feature of chiral perturbation theory in which the imaginary part of
the amplitude is not calculated as accurately as the real part, and
thus unitarity is only approximately satisfied at a given order
\cite{GM}. As will be discussed below the difference between the
unitary and one loop ChPT values of $\beta$ can be observed in 
future
experiments in which $\hbox{Im} E_{0+}$ will be measured directly.

\begin{figure}
\begin{center}
\epsfig{figure=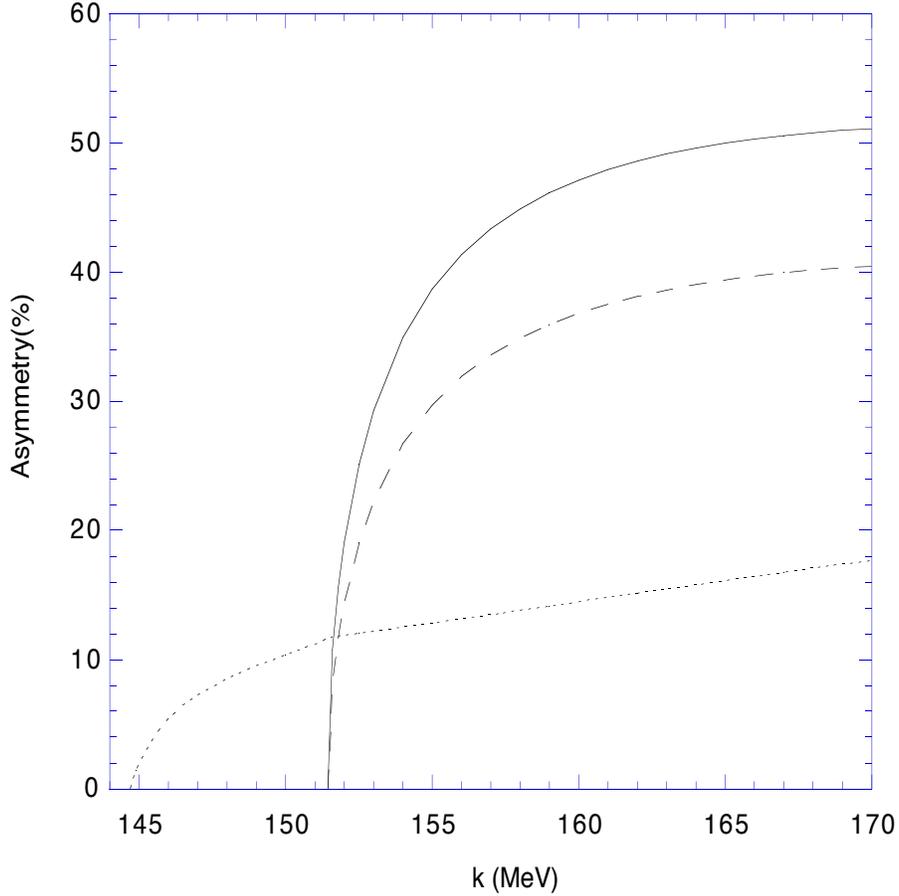,width=12cm,height=12cm}
\end{center}
\caption{The polarized photon ($\Sigma$) and polarized target (T)
asymmetries (in \%) at a $\pi^{0}$ center of mass angle of $90^{o}$
for the $\gamma p \rightarrow \pi^{0}p$ reaction versus photon 
energy
k.  The lower curve is the ChPT calculation \cite{BKM} for
$\Sigma$. The two curves for T which rise rapidly at the $\pi^{+}n$
threshold at 151.4 MeV are the ChPT calculation (dashed curve)
\cite{BKM} and the unitary calculation (solid curve).}\label{fig3}
\end{figure}

To accurately measure the magnitude of the unitary cusp and to 
exploit
the connection between electromagnetic pion production and low 
energy
$\pi N$ interactions, one must measure $\hbox{Im} E_{0+}$. In
photoproduction this requires experiments with polarized beams 
and/or
targets \cite{AB}.  For the sake of brevity the formulas connecting
the cross sections and polarization observables to the multipoles
\cite{multipoles} will not be quoted here. To briefly demonstrate the
power of polarized photo-pion experiments, two asymmetries are 
shown
in Fig.~\ref{fig3}: $\Sigma$ for linearly polarized photons with an
unpolarized target; and T for unpolarized photons but with a target
polarized normal to the reaction plane. The results presented in
Fig.~\ref{fig3} use the p wave predictions of chiral perturbation
theory \cite{BKM} and the unitary fit to $E_{0+}$ discussed above
\cite{Mainz}. $\Sigma$ is primarily sensitive to the p wave 
multipoles
and since the unitary fit has essentially the same p wave multipoles
as ChPT, the curves for ChPT and the unitary fit are almost
identical. By contrast, T is sensitive to a linear combination of p
wave multipoles times $\hbox{Im} E_{0+}$, and shows its rapid rise
above the $\pi^{+}n$ threshold. For T, the large difference in
$\hbox{Im} E_{0+}$ between the unitary fit \cite {Mainz} and ChPT to
one loop \cite{BKM} should be straightforward to distinguish
experimentally. A proposed experiment with tagged photons using 
an
active, polarized proton target at Mainz estimates that $\beta$ can be
measured to $\simeq1$ to 2\% \cite{polp}. Similar results could be
obtained using a laser backscattering source \cite{Blaine, Duke} or
small angle electron scattering with polarized, internal, targets in a
storage ring facility \cite{Polite}.

There are several possible strategies to test isospin conservation
from measurements in the near threshold $\gamma N\rightarrow 
\pi N$
reaction.  A definitive demonstration of this effect would involve
precision measurements of the real and imaginary parts of $E_{0+}$ 
for
the unitary cusp contributions to the $\gamma n 
\rightarrow\pi^{0}n$
and $\gamma p \rightarrow \pi^{0}p$ reactions. Although the 
predicted
effect of $\simeq$ 2 to 3\% in $\beta$ is larger than the usual
$\simeq$1\% effect expected on the basis of electromagnetic effects,
it represents a serious experimental challenge to accomplish this
goal, particularly since free neutron targets are not available to
make this measurement. A simpler strategy would be to perform a
precision measurement of the unitary cusp for the $\gamma 
p\rightarrow
\pi^{0}p$ reaction, extract the value of $\beta$, and compare the
result of with the value obtained using unitarity and isospin
conservation quoted above. An equivalent strategy is to obtain
$a_{cex}(\pi^{+} n\rightarrow\pi^{0}p)$ from the measured value of
$\beta$ and compare it to the accurately measured value of
$a_{cex}(\pi^{-}p\rightarrow\pi^{0}n)$ \cite{PSI}, obtained from the
width of the 1s state in pionic hydrogen, to see if there is any
deviation from the pure isospin prediction that these values are 
equal
and opposite \cite{sign}. It is important to note that there is a PSI
proposal \cite{PSI} to improve the pionic hydrogen measurement so 
that
the error in $a_{cex}(\pi^{-}p\rightarrow\pi^{0}n)$ will be reduced to
$\simeq$ 1\%.

It is therefore of interest to discuss the information that might be
obtained from measurements of the s wave $\pi N$ scattering 
lengths,
$a(\pi N)$, from photopion reactions with the values that have been
obtained using conventional pion beams. The values for $a(\pi N)$ 
for
several physical channels are presented in Table~\ref{tab1} (the
appropriate isospin relations \cite{Hohler} were used when
required). The most accurate and direct measurements come from 
pionic
hydrogen experiments \cite{PSI}. In order to test isospin symmetry 
the
PSI group has also measured pionic deuterium in order to get at the
$\pi^{-} n$ scattering length \cite{PSI}. Unfortunately the two body
corrections in deuterium are large and add significantly to the error
which makes the isospin conservation test uncertain at the required
level of accuracy, so that these values are not quoted here. The
results of the two most recent empirical analyses of the $\pi N$ data
for pion kinetic energies greater than $\simeq$ 30 MeV, extrapolated
to threshold, are also presented \cite{Matsinos,VPI} in
Table~\ref{tab1}. In the SAID study the scattering
lengths come from a dispersive analysis, the errors are hard to
ascertain and are not quoted \cite{VPI}. The chiral perturbation
theory results \cite{Meissner} are consistent with experiment. If
isospin symmetry is exact there are two independent scattering 
lengths
in the I =1/2, 3/2 states, and six possible physical $\pi$N elastic
scattering and charge exchange reactions, so that there are many
possible tests if three or more channels can be measured
precisely. This is precisely the gap that can be filled by measurement
of the final state $\pi$N interactions in photopion production, e.g.
$\pi^{0}p$ or $\pi^{+}n \rightarrow \pi^{0}p$ in the $\gamma p
\rightarrow \pi^{0} p$ reaction.

\begin{table}
\caption{S-wave $\pi N$ scattering lengths $a(\pi N)$
for several channels in units of $10^{-2}/m_{\pi}$ (N = n or p)}
\label{tab1}
\begin{center}
{\small
\begin{tabular}{||c|c|c|c|c|c||}\hline\hline
 & & & & & \\
       &  Pionic  &   Matsinos    &  SAID    &     ChPT      &Estimated\\
Channel&  Atoms   &\cite{Matsinos}& (Sp98)   &\cite{Meissner}& 
isospin \\
       &\cite{PSI}&               &\cite{VPI}&               
&breaking\cite{W1,vanKolk}\\
 & & & & & \\ \hline
 & & & & & \\
$\pi^-p$ & 8.83(0.08) & 8.14(0.10) & 8.83 & 8.70(0.86) &  $\simeq 0 
$\\
$\pi^-p\rightarrow \pi^{0}n$
         & --13.01(0.59) & --10.93(0.08) & --12.5 & --12.45(0.75)
         & $\simeq$ 2 to 3 \% \\
$\pi^{0} N$ &  & 0.41(0.09) &  0.0 & --0.1(0.7) & $\simeq$ 30\% \\
 & & & & & \\ \hline\hline
\end{tabular} }
\end{center}
\end{table}

Another test of isospin conservation has been made for medium 
energy
$\pi$N scattering (pion kinetic energy from 30 to 100
MeV) \cite{Matsinos,Gibbs}. In both cases the scattering amplitudes 
in
the I = 1/2 and 3/2 states were obtained from the data for 
$\pi^{\pm}
p$ elastic scattering. From these, the prediction for the $\pi^{-}p
\rightarrow \pi^{0} n$ charge exchange amplitude made on the
assumption of isospin conservation was compared to the empirical
charge exchange amplitude. An isospin violation at the $\simeq 7\%$
level has been found by both analyses \cite{Matsinos,Gibbs}, 
primarily
in the s wave amplitude. These analyses depend on the quality of the
data and on the accuracy of the Coulomb corrections, which have 
been
criticized as being inconsistent with the strong interaction
calculations that were employed \cite{Meissner}. If isospin is indeed
violated at the 7\% level, it is not clear how to relate this to the
isospin breaking predictions for the s wave scattering
length \cite{W1,vanKolk} which is applicable at lower energies. What 
is
required is an extension of the isospin breaking calculations to
medium energies. A test of the existing predictions
requires experiments at lower energies which would measure the s 
wave
scattering lengths.

In addition to the measurement of $\hbox{Im} E_{0+}$ above the
$\pi^{+}n$ threshold discussed above, one could contemplate the 
more
difficult measurement when only the $\pi^{0}p$ channel is open. As 
was
discussed previously, from a measurement of both the $\hbox{Re}$ 
and
$\hbox{Im}$ parts of $E_{0+}$ (or $L_{0+}$) one obtains $\delta_{0}$,
for which there is a predicted $\simeq 30\%$ isospin breaking
effect. However, because of the small expected size of $\delta_{0}
\simeq 1^{o}$ in the energy region below the $\pi^{+}n$ threshold,
this is a very difficult task. It may be at the limits of feasibility
using either a laser backscattering source, where an intense photon
flux is concentrated in a small energy interval \cite{Blaine, Duke},
or with internal target, small angle scattering with polarized
internal targets \cite{Polite}.

In conclusion, the connection between the $\gamma^{* }p\rightarrow
\pi^{0} p$ reaction and $\pi N $ scattering has been presented in a
rigorous, model-independent way, which is unitary and time reversal
invariant, and where the isospin breaking due to the threshold
difference between the $\pi^{0} p$ and $\pi^{+} n$ channels, and the
mass difference between the up and down quarks, is taken into
account. This leads to a predicted unitary cusp due to the two step
$\gamma p \rightarrow \pi^{+} n \rightarrow \pi^{0} p $ charge
exchange reaction. The magnitude of the unitary cusp is given by
$\beta\equiv E_{0+}(\gamma p \rightarrow\pi^{+}n) \cdot
a_{cex}(\pi^{+}n\rightarrow\pi^{0}p)$. This unitary cusp has been
recently observed in photoproduction experiments \cite{Mainz,Sask}.
It has been shown for the first time that there is a dynamical isospin
breaking effect in the value of $\beta$, due to the mass difference of
the up and down quarks, in electromagnetic pion production.  This is
linked by unitarity and time reversal invariance to a predicted quark
mass effect in $\pi^{0} N $ scattering and pion charge
exchange \cite{W1,Meissner,vanKolk}. At present there are accurate
measurements for $a_{\pi^{-}p}$ and $a_{cex}(\pi^{- }p\rightarrow
\pi^{0}n)$ from pionic atoms performed at PSI \cite{PSI}. To check
isospin conservation requires at least one more precision 
measurement
in another charge channel. This is more readily performed in
electromagnetic meson production for two reasons: 1) in order to
accurately measure the s wave scattering length, it is important to
work at very low energies (e.g. $\leq$ 10 MeV) at which $\pi N$
experiments are hard to perform since the low energy charged pions
decay and also since $\pi^{0}$ beams can not be made at any energy;
and 2) in electromagnetic pion production one can access charge 
states
that cannot be reached with conventional pion beams (this is 
important
for isospin checks). It is shown that photoproduction experiments 
with
polarized targets can lead to a precise measurement of the small but
interesting isospin violating effects. In addition to the polarization
observables in photoproduction discussed above, similar information
can be obtained in threshold $\pi^{0}$ electroproduction from the
combination of the transverse- longitudinal (TL and TL') structure
functions \cite{multipoles}. This possibility will be discussed in a
future publication. Finally we stress that the important program of
precisely measuring the $\pi N$ and photopion reactions that are
required to test the predicted isospin violation, is difficult but
feasible.

\section{Acknowledgments}
 I would like to thank F.E.Low for his collaboration 
in deriving Eq. 2. For many fruitful discussions I would like to thank 
U.G.~Mei{\ss}ner, N.Kaiser, B.Holstein, and U.B.vanKolk. For critical 
help with the manuscript I would like to thank M.Distler and 
M.Pavan.


\begin{thebibliography}{99}
\bibitem{notation} There is a sign ambiguity in $|f_{cex}|$ which is
resolved by taking the sign from $\hbox{Re} f_{cex} \rightarrow
a_{cex} q_{c}$ near threshold.
\bibitem{units} The units of scattering lengths are $10^{-2}/m_{\pi}$
and for $E_{0+}$  and $\beta$ are $10^{-3}/m_{\pi}$.
\bibitem{sign} The sign depends on the isospin convention used and
can  be either positive \cite{W1,vanKolk} or negative \cite{BKM}. We
use here the latter convention.
\end{thebibliography}
\end{document}